# High-Rate Quantization for the Neyman-Pearson Detection of Hidden Markov Processes

Joffrey Villard*, Pascal Bianchi†, Éric Moulines† and Pablo Piantanida*

*SUPELEC, Telecom. Dpt., Gif-sur-Yvette, France

†LTCI Telecom Paristech, Paris, France

Email: {joffrey.villard,pablo.piantanida}@supelec.fr, {bianchi,moulines}@telecom-paristech.fr

**Abstract**

This paper investigates the decentralized detection of Hidden Markov Processes using the Neyman-Pearson test. We consider a network formed by a large number of distributed sensors. Sensors' observations are noisy snapshots of a Markov process to be detected. Each (real) observation is quantized on $\log_2(N)$ bits before being transmitted to a fusion center which makes the final decision. For any false alarm level, it is shown that the miss probability of the Neyman-Pearson test converges to zero exponentially as the number of sensors tends to infinity. The error exponent is provided using recent results on Hidden Markov Models. In order to obtain informative expressions of the error exponent as a function of the quantization rule, we further investigate the case where the number $N$ of quantization levels tends to infinity, following the approach developed in [1]. In this regime, we provide the quantization rule maximizing the error exponent. Illustration of our results is provided in the case of the detection of a Gauss-Markov signal in noise. In terms of error exponent, the proposed quantization rule significantly outperforms the one proposed by [1] for i.i.d. observations.

## I. INTRODUCTION

The design of powerful tests allowing to detect the presence of a stochastic signal using large Wireless Sensor Networks (WSN) is a crucial issue in a wide range of applications. In many practical contexts, each sensor of the network gathers information on its environment and forwards it to a distant Fusion Center (FC) which makes the final decision. Binary hypothesis tests are special cases where the FC has to decide between two possible hypotheses $H_0$ and $H_1$. Consider a network composed of $n+1$ sensors, denote by $Y_k$ the random variable representing the observation of the $k$th sensor ($k = 0 \ldots n$) and by $p_0(y_{0:n})$ (resp. $p_1(y_{0:n})$) the probability distribution of the global observation vector $Y_{0:n}$ under $H_0$ (resp.

The works of J. Villard are supported by DGA (French Armement Procurement Agency).

$H_1$). The task of the FC is then to decide between the following hypotheses:

$$H_0: Y_{0:n} \sim p_0(y_{0:n})$$
$$H_1: Y_{0:n} \sim p_1(y_{0:n}) .$$

For any test function, we refer to the *probability of false alarm* (resp. the *miss probability*) as the probability that the FC decides hypothesis $H_1$ (resp. $H_0$) under hypothesis $H_0$ (resp. $H_1$). The Neyman-Pearson test consists in rejecting the null hypothesis whenever the (normalized) log-likelihood ratio (LLR) $L_n$ defined by

$$L_n = \frac{1}{n} \log \frac{p_0(Y_{0:n})}{p_1(Y_{0:n})} \quad (1)$$

lies below a certain threshold. This threshold is set in such a way that the probability of false alarm is no larger than a certain *level* $\alpha \in (0,1)$ [2]. The associated miss probability $\beta_n(\alpha)$ is the key metric to characterize the performance of the detection procedure. Unfortunately, no simple closed form expression is available in general. However, under some assumptions on the law of the observations, the following lemma due to Chen [3] shows that the miss probability behaves as $\beta_n(\alpha) \simeq \exp(-nK)$ for large $n$, where $K$ is a certain positive constant, independent of $\alpha$, which we will refer to as the *error exponent*.

**Lemma 1** ([3]). *Assume that $L_n$ converges in probability to a certain constant $K > 0$ under $H_0$. Then, for any $\alpha \in (0,1)$,*

$$\lim_{n\to\infty} -\frac{1}{n} \log \beta_n(\alpha) = K .$$

The error exponent $K$ provides crucial information on the performance of the NP test when the number $n$ of sensors is large. By Lemma 1, the evaluation of $K$ simply reduces to the asymptotic analysis of the LLR $L_n$ under $H_0$. In this framework, a number of works in the literature derived and analyzed the error exponent for various observation models (see for instance [4], [5] and reference therein). However, most of these works assume that the FC has a perfect knowledge of the sensors' observations. Unfortunately, in a WSN, the amount of information forwarded by each sensor node to the FC is usually limited, due to imperfect links between nodes of the network. Therefore, a large numbers of papers have been devoted to the construction and the analysis of *decentralized* detection schemes [1], [6]–[9]. In this framework, each sensor has the ability to compress/quantize its observation before transmission to the FC, with the aim of decreasing the information transport burden to be supported by the network. For instance, [7], [9] investigate the detection of deterministic signal in noise, while [1], [6], [8] study the case where all observations are independent and identically distributed (i.i.d.).

Few is known when the observations are correlated. In [10], the authors explore the effect of node density on the error exponent for the detection of correlated signals in Gaussian noise, assuming that the sensors communicate with the FC through Gaussian additive channels. In this paper, we investigate the more general case where the observations follow a Hidden Markov Model (HMM). More precisely, the source signal to be detected is supposed to be a Markov chain. Sensors observations are distorted (noisy) versions of the latter source signal. Each observation is quantized on $log_2(N)$ bits before being transmitted to the FC. Our aim is to evaluate the impact of quantization on the error exponent, and to characterize relevant quantizers allowing to maximize the error exponent.

The rest of the paper is organized as follows. In Section II, we describe the observation model and evaluate the error exponent in the ideal case where the FC has perfect access to the observations. In Section III, we introduce *high-rate* quantizers, and evaluate the degradation on the error exponent when the decision is made using *quantized* observations instead of the ideal ones. We determine relevant quantization strategies allowing to reduce this degradation. Section IV illustrates our results for the detection of a Gauss-Markov signal in noise.

## II. Detection of Hidden Markov Models

### A. Model of Observation

In this section, we describe the probabilistic model underlying the observations. We assume that the observed time series $Y_k$ follows a Hidden Markov Model under both hypotheses $H_0$ and $H_1$, with different transition kernels. Our hypothesis testing problem shall thus reduce to the question: Which kernel underlies the observation process ?

To be more specific, let $(X, \mathcal{X})$ and $(Y, \mathcal{Y})$ be two measurable spaces. Consider two transition kernels $Q_0$ and $Q_1$ defined on $X \times \mathcal{X}$. We make the assumption that both kernels are *positive* and denote by $\nu_i$ the invariant measure of $Q_i$ [11]. Let $G$ be a transition kernel on $X \times \mathcal{Y}$. Now consider two probability measures $P_0$ and $P_1$ on a relevant measurable space $(\Omega, \mathcal{F})$ and a stochastic process $(X_k, Y_k)_{k \in \mathbb{Z}}$ such that for each $i \in \{0, 1\}$, the following holds true under probability measure $P_i$:

- $(X_k, Y_k)_{k \in \mathbb{Z}}$ is stationary.
- $(X_k, Y_k)_{k \geq 0}$ is a Markov chain with kernel $T_i$ defined for each $(x, y) \in X \times Y$ and each $C \in \mathcal{X} \otimes \mathcal{Y}$ by

$$T_i[(x,y), C] = \int \int_C Q_i(x, dx') G(x', dy') . \qquad (2)$$

- $X_0$ has distribution $\nu_i$.

In particular, definition (2) implies that the *state* $X_k$ is a Markov chain with transition kernel $Q_i$ under $P_i$. The observation $Y_k$ is such that

$$P_i[Y_k \in B | X_k] = G(X_k, B) , \ \forall B \in \mathcal{Y} . \qquad (3)$$

The aim is to decide between probability measures $P_0$ and $P_1$ based on the observation of $Y_{0:n}$, the state sequence $X_{0:n}$ being unobserved. Note that Equation (3) implies that the distribution of the observation $Y_k$ conditionnally to the state $X_k$ is identical under both $P_0$ and $P_1$. Roughly speaking, one can think of $Y_k$ as a noisy version of a process $X_k$ to be detected, where the distortion (typically a measurement noise) does not depend on the hypothesis. We make the following assumptions.

**A1**. *There exist two probability measures $\lambda$ and $\mu$ on $(X, \mathcal{X})$ and $(Y, \mathcal{Y})$ respectively s.t. for each $x \in X$ and each $i \in \{0,1\}$, $Q_i(x,.)$ admits a density $x' \mapsto q_i(x,x')$ w.r.t. $\lambda$ and $G(x,.)$ admits a density $y \mapsto g(x,y)$ w.r.t. $\mu$.*

**A2**. *There exist two real numbers $\sigma^-$, $\sigma^+$ s.t., for each $i \in \{0,1\}$ and each $x, x' \in X$, $0 < \sigma^- \leq q_i(x, x') \leq \sigma^+$.*

**A3**. $0 < \int g(x,y)\lambda(dx) < \infty$ *for each $y \in Y$.*

**A4**. $\sup_{x,y} g(x,y) < \infty$ .

**A5**. $E_0 \left| \log \int g(x, Y_0)\lambda(dx) \right| < \infty.$

We mention that **A2** implies some restrictions on the state model. For instance, this assumption generally excludes state models with unbounded support such as the Gaussian autoregressive model considered in [4]. Note that milder assumptions, under which most of the theoretical results used in this paper still holds true, can be found in [12]. Such an extension is however out of the scope of this paper.

For each $i \in \{0,1\}$ and each integers $k \leq \ell$, denote by $p_i(y_{k:\ell})$ the density of measure $P_i[Y_{k:\ell} \in .]$ w.r.t. to measure $\mu^{\otimes(\ell-k+1)}$. Denote by $p_i(y_\ell | Y_{k:\ell-1})$ the density of $P_i[Y_\ell \in . | Y_{k:\ell-1}]$ w.r.t. $\mu$.

*B. Error Exponent for Unquantized Observations*

In this paragraph, we derive the error exponent associated with the NP test in the ideal case where the FC is supposed to have perfect access to the sensors' observations $Y_{0:n}$. From Lemma 1, the derivation of the error exponent $K$ reduces to the study of the LLR $L_n$ under $P_0$. The following result can be easily proved from [12].

**Theorem 1** ([12]). *Under Assumptions* **A1**–**5**, *the following holds true.*

*i) For each $i \in \{0,1\}$ and each $k \geq 0$, the following limit exists with probability one under $P_0$:*

$$\mathcal{L}_i(Y_{-\infty:k}) = \lim_{m \to \infty} \log p_i(Y_k | Y_{-m:k-1}) . \quad (4)$$

*Moreover, $E_0 |\mathcal{L}_i(Y_{-\infty:k})| < \infty$.*

*ii) Under $P_0$, the log-likelihood ratio $L_n$ defined by (1) converges a.s. as $n \to \infty$ to the constant $K$ defined by*

$$K = E_0 [\mathcal{L}_0(Y_{-\infty:0})] - E_0 [\mathcal{L}_1(Y_{-\infty:0})] .$$

**Elements of the Proof.**

Here we only recall the general ideas and refer to [12] for the detailed motivation of the forthcoming statements. The first step of the proof consists in rewriting the LLR $L_n$ using the so-called chain rule. Equation (1) becomes

$$L_n = \frac{1}{n} \sum_{k=0}^{n} \log p_0(Y_k|Y_{0:k-1}) - \frac{1}{n} \sum_{k=0}^{n} \log p_1(Y_k|Y_{0:k-1}) . \quad (5)$$

Thus, the asymptotic analysis of $L_n$ as $n \to \infty$ reduces to the separate study of each of the two terms of the rhs of the above equation. Focus for instance on the second term. The main step of the proof is to establish the following *forgetting property* of the past observations, for each $k \geq 0$ and $m' \geq m \geq 1$:

$$|\log p_1(Y_k|Y_{-m:k-1}) - \log p_1(Y_k|Y_{-m':k-1})| \leq \frac{2}{1-\rho} \rho^{k+m-1}, \quad (6)$$

where we defined $\rho = \sigma^-/\sigma^+ < 1$. The core of the proof of the above inequality is essentially related to the *mixing condition* **A2** which implies a Dobrushin's contraction condition on the forward smoothing kernels. From inequality (6), $(\log p_1(Y_k|Y_{-m:k-1}))_{m \geq -k}$ is therefore a Cauchy sequence. Denote its limit by $\mathcal{L}_1(Y_{-\infty:k})$. Plugging (4) into (6) and using the triangular inequality, it is straightforward to show that

$$\left| \frac{1}{n} \sum_{k=0}^{n} \log p_1(Y_k|Y_{0:k-1}) - \frac{1}{n} \sum_{k=0}^{n} \mathcal{L}_1(Y_{-\infty:k}) \right| \xrightarrow[n \to \infty]{P_0\text{-a.s.}} 0 \ .$$

Assumptions **A4** and **A5** ensure that $E_0|\mathcal{L}_1(Y_{-\infty:0})| < \infty$. As process $Y_k$ is stationary under $P_0$, the ergodic theorem along with the latter equation leads to

$$\frac{1}{n} \sum_{k=0}^{n} \log p_1(Y_k|Y_{0:k-1}) \xrightarrow[n \to \infty]{P_0\text{-a.s.}} K_1 = E_0\left[\mathcal{L}_1(Y_{-\infty:0})\right] \ .$$

By the same token, the first term of the rhs of Equation (5) converges towards $K_0 = E_0[\mathcal{L}_0(Y_{-\infty:0})]$. Therefore, $L_n$ converges to $K_0 - K_1$. This proves Theorem 1.

### III. MAIN RESULTS

From now on, we assume that each sensor's observation belongs to an interval of the form $Y = [a, b]$ for some $a < b$. We furthermore assume that $\mu$ is the Lebesgue measure on $Y$ normalized in such a way that $\mu(Y) = 1$.

*A. Definitions*

We now investigate the case where the final decision is made from quantized observations. For a given integer $N \geq 1$, we define an $N$-point *quantizer* as a triplet $\mathcal{Q}_N = (\mathcal{S}_N, \Xi_N, \xi_N)$ where $\mathcal{S}_N$ is a set of $N$ intervals $S_{N,0}, \ldots, S_{N,N-1}$ which form a partition of $Y$, where $\Xi_N = \{\xi_{N,0}, \ldots, \xi_{N,N-1}\}$ is an arbitrary set of distinct elements and where $\xi_N : Y \to \Xi_N$ is a function s.t. $\xi_N(y) = \xi_{N,j}$ whenever $y \in S_{N,j}$. We will refer to each interval $S_{N,j}$ as a *cell* of the quantizer. If $Y_k$ represents the value of the $k$th sensor's observation, we denote by $Z_{N,k} = \xi_N(Y_k)$ the quantized observation on $\log_2(N)$ bits. From now on, we assume that only the quantized observations $Z_{N,0:n} = (Z_{N,0} \ldots Z_{N,n})$ are available at the FC. Following the terminology of [1], we introduce some useful characteristics of a given quantizer $\mathcal{Q}_N$:

- The *length* of cell $j$ is defined by $\ell_{N,j} = \int_{S_{N,j}} dy$.
- The *specific point density* in cell $j$ is defined by $\zeta_{N,j} = \frac{1}{N\ell_{N,j}}$. For convenience, we also define function $\zeta_N$ on $Y$ by $\zeta_N(y) = \zeta_{N,j}$ whenever $y \in S_{N,j}$.

*B. Neyman-Pearson Test on Quantized Observations*

We derive the NP procedure for testing $P_0$ *vs* $P_1$. Consider the following weighted counting measure of the $\xi_{N,j}$'s:

$$\mu_N = \sum_{j=0}^{N-1} \frac{\ell_{N,j}}{b-a}\, \delta_{\xi_{N,j}} \ .$$

Clearly, the joint distribution $P_i[Z_{N,0:n} \in \,.\,]$ of the quantized observation vector is absolutely continuous w.r.t. $\mu_N^{\otimes(n+1)}$. We denote its density by $p_{i,N}(z_{0:n})$. The NP test consists in rejecting the null hypothesis for small values of the LLR $L_{n,N}$ associated with the quantized observations:

$$L_{n,N} = \frac{1}{n} \log \frac{p_{0,N}(Z_{N,0:n})}{p_{1,N}(Z_{N,0:n})} \ .$$

We now study the error exponent of the latter test. Note that the event $[Z_{N,k} = \xi_{N,j}]$ is equivalent to the event that $Y_k$ falls into cell $S_{N,j}$. Using (3), we thus obtain for each $i \in \{0,1\}$

$$P_i[Z_{N,k} = \xi_{N,j} \mid X_k] = G(X_k, S_{N,j}) \ .$$

As a consequence, process $(X_k, Z_{N,k})_k$ still forms a HMM, where the transition kernel $G_N$ which links the state $X_k$ to the quantized observation $Z_{N,k}$ is given by $G_N(x, \xi_{N,j}) = G(x, S_{N,j})$ for each $j \in \{1, \ldots, N\}$. We remark that $G_N$ admits the following density $g_N$ w.r.t. $\mu_N$:

$$g_N(x, \xi_{N,j}) = \frac{1}{\ell_{N,j}} \int_{S_{N,j}} g(x,y)\, dy \ . \tag{7}$$

If Assumptions **A1**–5 hold, then it is straightforward to show that the new (quantized) HMM $(X_k, Z_{N,k})_k$ still verifies conditions of same kind. This ensures that Theorem 1 still applies to the distorted LLR $L_{n,N}$. Thus, we obtain the following corollary. Define $z \mapsto p_{i,N}(z|Z_{N,-m:-1})$ as the derivative of $P_i[Z_{N,0} \in .|Z_{N,-m:-1}]$ w.r.t. $\mu_N$.

**Corollary 1.** *Under $P_0$, $L_{n,N}$ converges a.s. as $n \to \infty$ towards the constant $K_N$ defined by*

$$K_N = E_0\left[\mathcal{L}_{0,N}(Z_{N,-\infty:0})\right] - E_0\left[\mathcal{L}_{1,N}(Z_{N,-\infty:0})\right] \ , \tag{8}$$

*where for each $i \in \{0,1\}$, $\mathcal{L}_{i,N}(Z_{N,-\infty:0})$ is the almost sure limit under $P_0$ of $\log p_{i,N}(Z_{N,0}|Z_{N,-m:-1})$ as $m \to \infty$.*

The above corollary provides the error exponent $K_N$ associated with the NP test on quantized observations. A natural question is: How does the choice of the quantizer $\mathcal{Q}_N$ affect the detection performance ? Unfortunately, the error exponent depends on the cells $S_{N,j}$ in a rather involved way which does not immediately allow to evaluate the impact of the quantizer. In the sequel, we follow the approach of [1], [13] and focus on the case where the order $N$ of the quantizer tends to infinity. We refer to such quantizers as *high-rate* quantizers. This approach leads to a convenient and informative asymptotic expression of $K_N$, which can be easily maximized as a function of the quantizer.

*C. High-Rate Quantizers*

Consider a family of quantizers $(\mathcal{Q}_N)_{N\geq 1}$. We make the following assumption.

**A6**. *As $N \to \infty$, $\zeta_N$ converges uniformly to a certain continuous function $\zeta$ s.t. $\inf_{y\in Y} \zeta(y) > 0$.*

We will refer to $\zeta$ as the *model point density* of the family $(\mathcal{Q}_N)_{N\geq 1}$. For each $y \in Y$, $\zeta(y)$ can be interpreted as the asymptotic density of cells in the neighborhood of $y$. Intuitively, high-rate quantizers should be constructed in such a way that $\zeta(y)$ is large at those points $y$ for which a fine quantization is essential to discriminate the two hypotheses. Theorem 2 below provides a more rigorous formulation of this intuition. We need further assumptions.

**A7**. *For each $x$, function $y \mapsto g(x,y)$ is of class $C_3$ on $Y$.*

**A8**. $\inf_{x,y} g(x,y) > 0$ *and* $\sup_{x,y} \left|\frac{\partial^3 g}{\partial y^3}(x,y)\right| < \infty$.

We now provide the main result. Recall that $p_0(y)$ is the pdf of $Y_0$ under $P_0$.

**Theorem 2.** *Under Assumptions* **A1–8**, *the following statements hold true.*

*i) The following limit exists with probability one under $P_0$:*

$$\ell(Y_{-\infty:\infty}) \stackrel{def}{=} \lim_{k\to\infty} \lim_{m\to\infty} \frac{\partial \log \frac{p_0}{p_1}}{\partial y_0}(Y_{-m:k}) \ .$$

*Moreover, $|\ell(Y_{-\infty:\infty})| < C$ for some constant $C$.*

*ii) As $N$ tends to infinity, $N^2(K - K_N)$ converges to a constant $D_\zeta$ given by*

$$D_\zeta = \frac{1}{24} \int \frac{p_0(y)F(y)}{\zeta(y)^2} \mu(dy) \ , \qquad (9)$$

*where $F(y) = E_0\left[\ell(Y_{-\infty:\infty})^2 \,\Big|\, Y_0 = y\right]$.*

*iii) Moreover,*

$$D_\zeta \geq \frac{(b-a)^2}{24} \left(\int [p_0(y)F(y)]^{1/3} \mu(dy)\right)^3 \ ,$$

*where equality holds when*

$$\zeta(y) = \frac{[p_0(y)F(y)]^{1/3}}{\int [p_0(s)F(s)]^{1/3} ds} \ . \qquad (10)$$

Theorem 2 states that when the order of the quantizer tends to infinity, the error exponent $K_N$ associated with the NP test converges at speed $\frac{1}{N^2}$ to the error exponent $K$ that one would have obtained in the absence of quantization. Roughly speaking, if $\beta_{n,N}(\alpha)$ represents the miss probability of the NP test of level $\alpha$, the approximation

$$\beta_{n,N}(\alpha) \simeq e^{-n\left(K - \frac{D_\zeta}{N^2}\right)}$$

is valid when both the number $n$ of sensors and the order $N$ of quantization are large, but $n \gg N$. The loss in error exponent depends on the quantizer family only via its model point density $\zeta$. Expression (10) provides the optimal choice of $\zeta$ *i.e.*, the model point density which minimizes the loss in error exponent.

Note that this expression is quite similar to Bennett's one [14] which gives the MSE-optimal model point density (in this case $F(y) = 2$ on $Y$).

In the particular case where the observations are i.i.d., our expression of the loss $D_\zeta$ is consistent with the one obtained by Gupta and Hero (see Equation (20) in [1]).

**Insights on the Proof.**

Due to the lack of space, we only provide some of the basic ideas underlying the proof of Theorem 2. A rigorous proof will be provided in an extended version of this paper.

Since the error exponent $K_N$ does not depend on the particular choice of the quantization alphabet $\Xi_N$, we assume that each $\xi_{N,j}$ coincides with the center of cell $S_{N,j}$. We separately study each term $K_{i,N} = E_0\left[\mathcal{L}_{i,N}(Z_{N,-\infty:0})\right]$ that appears in Equation (8). First focus on $K_{1,N}$ and define $m = m(N)$ a sequence of integers such that $\frac{m}{\log N} \to \infty$ and $\frac{m}{N} \to 0$ as $N \to \infty$. We write

$$K_{1,N} - K_1 = E_0\left[\mathcal{L}_{i,N}(Z_{N,-\infty:0}) - \mathcal{L}_1(Y_{-\infty:0})\right]$$

$$= T_N + \delta_N \ ,$$

where $T_N = E_0\left[\log p_{1,N}(Z_{N,0}|Z_{N,-m:-1}) - \log p_1(Y_0|Y_{-m:-1})\right]$ and $\delta_N$ is the remainder term. The rhs of Equation (6) converges to zero exponentially as $m$ tends to infinity and the sequence $m(N)$ grows at a superlogarithmic rate. Consequently, the remainder $\delta_N$ is a little-o of $\frac{1}{N^2}$. The study of $T_N$ is based on the following expansions:

$$\log p_{1,N}(Z_{N,0}|Z_{N,-m:-1}) =$$
$$\log p_{1,N}(Z_{N,-m:0}) - \log p_{1,N}(Z_{N,-m:-1}) \ , \quad (11)$$

$$p_{1,N}(z_{-m:u}) =$$
$$\int \ldots \int \nu_1(dx_{-m}) \prod_{i=-m+1}^{u} Q_1(x_{i-1}, dx_i) \prod_{i=-m}^{u} g_N(x_i, z_i) \ . \quad (12)$$

Plugging Taylor-Lagrange expansion of $y \mapsto g(x,y)$ at point $\xi_{N,j}$ in (7) leads to the following approximate:

$$g_N(x, \xi_{N,j}) \approx g(x, \xi_{N,j}) + \frac{\ell_{N,j}^2}{24} \frac{\partial^2 g}{\partial y^2}(x, \xi_{N,j}) \ . \quad (13)$$

Note that the second term of the rhs of the above equation vanishes as $N \to \infty$, due to the fact that the cell lengths $\ell_{N,j}$ skrink. We now plug the above expansion into (12). The dominant term is nothing else than $p_1(z_{-m:u})$ (only keep term $g(x, \xi_{N,j})$ in (13)). Therefore, (11) is approximately equal to

$$\log p_1(Z_{N,0}|Z_{N,-m:-1}) + A_N(Z_{N,m:0})$$

where $A_N(Z_{N,m:0})$ contains the dominated terms. Therefore, the quantity of interest $T_N$ can be written as $T_N \simeq \mathcal{A}_N + \mathcal{B}_N$ where $\mathcal{A}_N = E_0\left[A_N(Z_{N,m:0})\right]$ and $\mathcal{B}_N = E_0\left[\log p_1(Z_{N,0}|Z_{N,-m:-1}) - \log p_1(Y_0|Y_{-m:-1})\right]$. After some algebra, and using recent results on HMM [12], we prove that $N^2 \mathcal{A}_N$ and $N^2 \mathcal{B}_N$ respectively converge to some constants $c_A$ and $c_B$ as $N \to \infty$. Thus, $N^2(K_{1,N} - K_1)$ converges to the sum $c_A + c_B$.

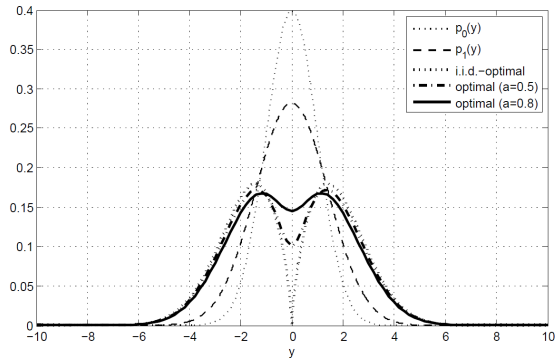

Fig. 1. Probability and model point densities ($\sigma = 1$)

Evaluating these constants, and proceeding in the same way for the study of $N^2(K_{0,N} - K_0)$, we obtain the first and second points of Theorem 2. From Holder's inequality, it is straightforward to prove the third point.

## IV. NUMERICAL RESULTS AND DISCUSSION

We consider the situation of detecting a Gaussian first-order autoregressive process (AR-1) embedded in noise [4]:

$$\begin{aligned} H_0 &: Y_k = W_k \\ H_1 &: Y_k = X_k + W_k \end{aligned} \quad (14)$$

where $X_k$ is a stationary Gaussian AR-1 process:

$$X_k = a X_{k-1} + \sqrt{1-a^2}\, U_k\,, \quad (15)$$

where $a \in (0,1)$ is the correlation coefficient, $U_k \stackrel{i.i.d}{\sim} \mathcal{N}(0,1)$ is the innovation process and $W_k \stackrel{i.i.d}{\sim} \mathcal{N}(0,\sigma^2)$ is the observation noise. We mention that in this case, all densities have infinite support so that, strictly speaking, the assumptions made in this paper are not satisfied. Nevertheless, the above model can be slightly modified to be consistent with our assumptions. For instance, in order that the transition kernel of (15) strictly fits Assumptions **A1-A2**, it is sufficient to replace the distribution $\mathcal{N}(0,1)$ of $U_k$ with the corresponding truncated distribution on an arbitrarily large support. In order to simplify the presentation, we do not go into more details and keep model (14)-(15) with slight abuse. We compare different quantization strategies in terms of the error exponent loss $D_\zeta$:

- optimal quantization: the model point density is given by Equation (10),
- i.i.d.-optimal quantization: the model point density is drawn as if observations were i.i.d. (see [1]) *i.e.*, we only use the marginal pdf of the observation to design the quantization rule,
- uniform quantization: all cells have the same size (in the interval $[-10\sigma, 10\sigma]$).

Figure 1 represents the model point densities for different values of parameter $a$. As the value of $a$ has no impact on the marginal distribution of the observations, the density associated to the i.i.d. quantizer is clearly independent of $a$. The latter density $\zeta_{iid}(x)$ is equal to zero at point $x = 0$, meaning that, if the observations were indeed i.i.d., fine quantization of the observations falling in the neighborhood of zero would be useless. On the opposite, the proposed quantization rule suggests to use a significant density of cells in the neighborhood of zero, especially when parameter $a$ is large. Figure 2 represents the error exponent loss $D_\zeta$ as a function of $a$ for $\sigma = 1$. As a matter of fact, when the model point density $\zeta = \zeta_{iid}$ is plugged into Equation (9), the integral diverges, due to the fact that $\zeta_{iid}(x)$ cancels at $x = 0$. Intuitively, this indicates that if the i.i.d. quantization rule is used to quantize the non-i.i.d. observations (14)-(15), the quantized error exponent does no longer converge to the perfect error exponent at speed $N^2$. Here, the proposed point density clearly outperforms the i.i.d. one in terms of asymptotic error exponent loss.

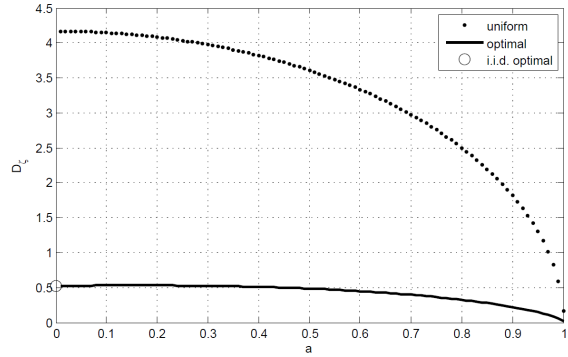

Fig. 2. Error exponent loss $D_\zeta$ for different quantization strategies ($\sigma = 1$)

## ACKNOWLEDGMENT

The authors would like to thank Prof. Alfred O. Hero for fruitful discussions.

## REFERENCES

[1] R. Gupta and A.O. Hero. High-rate vector quantization for detection. *IEEE Trans. Inf. Theory*, vol. 49, no. 8, Aug. 2003.

[2] E.L. Lehmann. *Testing Statistical Hypotheses, 2nd Edition*. Springer Texts in Statistics, 1997.

[3] P.-N. Chen. General formulas for the neyman-pearson type-ii error exponent subject to fixed and exponential type-i error bounds. *IEEE Trans. Inf. Theory*, vol. 42, no. 1, Jan. 1996.

[4] Y. Sung, L. Tong, and H.V. Poor. Neyman-pearson detection of gauss-markov signals in noise : closed-form error exponents and properties. *IEEE Trans. Inf. Theory*, vol. 52, no. 4, Apr. 2006.

[5] W. Hachem, E. Moulines, J. Najim, and F. Roueff. On the error exponents for detecting randomly sampled noisy diffusion processes. In *Proc. ICASSP*, Taipei, Taiwan, 2009.

[6] K. Liu and A.M. Sayeed. Type-based decentralized detection in wireless sensor networks. *IEEE Trans. Signal Process.*, vol. 55, no. 5, May 2007.

[7] J.-J. Xiao and Z.-Q. Luo. Universal decentralized detection in a bandwidth-constrained sensor network. *IEEE Trans. Signal Process.*, vol. 53, no. 8, Aug. 2005.

[8] J.N. Tsitsiklis. Decentralized detection by a large number of sensors. *Math. of control, signals and systems*, vol. 1, no. 2, 1988.

[9] P. Willett, P.F. Swaszek, and R.S. Blum. The good, bad, and ugly: Distributed detection of a known signal in dependent gaussian noise. *IEEE Trans. Signal Process.*, vol. 48, no. 12, Dec. 2000.

[10] J.-F. Chamberland and V.V. Veeravalli. How dense should a sensor network be for detection with correlated observations? *IEEE Trans. Inf. Theory*, 52(11):5099–5106, 2006.

[11] S.P. Meyn and R.L. Tweedie. *Markov Chains and Stochastic Stability*. Springer-Verlag, 1993.

[12] R. Douc, E. Moulines, and T. Ryden. Asymptotic properties of the maximum likelihood estimator in autoregressive models with markov regime. *The Annals of Statistics*, vol. 32, no. 5, 2004.

[13] A. Gersho and R.M. Gray. *Vector quantization and signal compression*. Kluwer, 1992.

[14] W.R. Bennett. Spectra of quantized signals. *BSTJ*, vol. 27, Jul. 1948.